\documentclass[jgrga]{agutex}

  \usepackage[dvips]{graphicx}
  \setkeys{Gin}{draft=false}
\authorrunninghead{L. H. YING and L. KAPLAN}

\titlerunninghead{ROGUE WAVE STATISTICS IN CNLS$^4$}

\authoraddr{L. H. Ying,
Department of Physics and Engineering Physics, Tulane University, New Orleans, Louisiana 70118, USA.
(lying@tulane.edu)}

\authoraddr{L. Kaplan,
Department of Physics and Engineering Physics, Tulane University, New Orleans, Louisiana 70118, USA.
(lkaplan@tulane.edu)}

\begin{document}

%
%

\title{Systematic Study of Rogue Wave Probability Distributions in a Fourth-Order Nonlinear Schr\"odinger Equation}
%
%

%
%



\authors{L. H. Ying \altaffilmark{1} and L. Kaplan \altaffilmark{1}}

\altaffiltext{1}{Department of Physics and Engineering Physics, Tulane University, New Orleans, Louisiana 70118, USA.}





%
%


\begin{abstract}
Nonlinear instability and refraction by ocean currents are both important mechanisms that go beyond the Rayleigh approximation and may be responsible for the formation of freak waves. In this paper, we quantitatively study nonlinear effects on the evolution of surface gravity waves on the ocean, to explore systematically the effects of various input parameters on the probability of freak wave formation. The fourth-order current-modified nonlinear Schr\"{o}dinger equation (CNLS$^4$) is employed to describe the wave evolution. By solving CNLS$^4$ numerically, we are able to obtain quantitative predictions for the wave height distribution as a function of key environmental conditions such as average steepness, angular spread, and frequency spread of the local sea state. Additionally, we explore the spatial dependence of the wave height distribution, associated with the buildup of nonlinear development.
\end{abstract}

%
%

%

\begin{article}

\section{Introduction}

Oceanic rouge waves, or freak waves, are surface gravity waves whose wave heights are extremely large compared to the typical wave in a given sea state. Freak waves have been reported throughout maritime history, as they are a hazard to mariners, cargo ships, and even to large cruise liners. However, serious scientific investigation commenced more recently, due to the availability of wave state measurement methods including satellite remote sensing, and the introduction of stochastic process theory to ocean wave forecasting~\citep{bates52,kinsman65}. 

The scientific community has studied the topic experimentally (in water tanks) ~\citep{onorato04} and~\citep{onorato06}, observationally (with satellite imaging, for example) ~\citep{dankert03} and~\citep{schulz04}, and theoretically~\citep{kharif03}. A review of recent progress in freak wave forecasting can be found in an article by~\citet{dysthe08}. Given the chaoticity of ocean wave dynamics, it is not possible to predict individual freak wave events. Instead, we study wave behavior probabilistically. The simplest theory assumes a random superposition of a large number of monochromatic waves with different frequencies and propagating directions, as in the Longuet-Higgins random seas model~\citep{longuet57}. By the central limit theorem, this model leads to a Rayleigh probability distribution of wave heights,
\begin{equation} \label{rayleigh}
  f_{\rm Rayleigh}(2H)=\frac{2H}{4\sigma^2}\exp\left(-\frac{(2H)^2}{8\sigma^2}\right) \,,
\end{equation}
where $2H$ is the wave height, $\sigma^2$ is the variance of the surface elevation, and $\sqrt{2\pi}\sigma$ is the mean wave height. For linear waves, the wave height $2H$ (crest to trough) equals twice the crest height due to a symmetry of the wave equation. This symmetry is broken for nonlinear waves, and throughout the paper, we focus on the wave height $2H$, where the crest height is $H$ to leading order.

Conventionally, freak waves are defined as having wave height $2H\geq 8.8\sigma$, or alternatively $2H\geq 2.2H_s$, where $H_s$ is the significant wave height~\citep{dean90}. The significant wave height $H_s$ was traditionally defined as the average of the tallest one-third of waves in a wave train, which comes to $H_s \approx 4.004\sigma$ when the height distribution is Rayleigh. Since the Rayleigh distribution (\ref{rayleigh}) serves as an excellent zeroth-order approximation for realistic wave height distributions in the ocean, $H_s=4\sigma$ may be used today as another definition of the significant wave height~\citep{forristall78}. Similarly, extreme freak waves are defined as having wave height $2H\geq 12.0\sigma$ or $2H \geq 3H_s$. According to the simple Rayleigh distribution (\ref{rayleigh}), the occurrence probabilities of a freak wave and an extreme freak wave are $6.3\times 10^{-5}$ and $1.5\times 10^{-8}$, respectively. Compared with the observational data~\citep{dankert03}, the stochastic Rayleigh model greatly underestimates the probability of freak waves. Recent advances in technology have allowed multiple wave tank experiments and field observations to be conducted, confirming the need for a more realistic theory to explain the results~\citep{forristall00, onorato04}. 

Nonlinear instability and refraction by ocean currents are both important mechanisms that go beyond the Rayleigh approximation and may be responsible for the formation of freak waves. 

The nonlinear instability, also known as the Benjamin-Feir instability in the one-dimensional case~\citep{benjamin67}, describes how a monochromatic wave can be unstable under a class of small perturbations. In the two-dimensional case, however, the spread in the wave vector happens not only in magnitude but also in wave direction. The nonlinear evolution involves both a nonlinear self-focusing effect at short times, which leads to an increase in the occurrence rate of extreme events, and also a dispersion in wave vector on larger time scales, which leads to a decrease in the probability of extreme events as the initial spectrum becomes less unidirectional~\citep{onorato02}. 

We usually use the steepness to describe the nonlinearity of a water wave. Steepness is defined as $\varepsilon=\overline{2H} k_0/2$, where $\overline{2H}$ is the mean wave height and $k_0$ is the peak wave number. The nonlinear evolution of a surface-gravity water wave is believed to be well described by the nonlinear Schr\"{o}dinger equation when the steepness is small, $ \overline{2H}  k_0\ll 1$, and the wave bandwidth is narrow, $|\Delta \vec{k}|/k_0 \sim \varepsilon \ll 1$. This is supported by Trulsen and Dysthe's analysis of the famous New Year's wave event, which occurred at the Draupner platform in the North Sea in 1995~\citep{trulsen97}. 

Clearly the exact evolution is always nonlinear to some extent, but the key is to introduce nonlinearities at the right moment and in an insightful and computable way. Realistic fully nonlinear computations wave by wave over large areas are very challenging, but initial attempts have been made to simulate the ocean surface using the full Euler equation both on large scales~\citep{tanaka01} and over smaller areas~\citep{gibson05, gibson07}. However, in this paper, we rely on series expansions, resulting in a sequence of approximations. Pioneering work using a third-order equation in $\varepsilon$ (the original Nonlinear Schr\"odinger Equation (NLS)) was performed by~\citet{zakharov68}, and then extended to fourth order by~\citet{dysthe79} (for water waves propagating in an ocean of infinite depth). In nature, the small steepness constraint is usually satisfied, since the steepness under storm conditions is normally less than 0.09~\citep{dysthe08}. However, the narrow bandwidth constraint is often not complied with in naturally occurring conditions. To overcome this problem, a modified nonlinear Schr\"{o}dinger equation (MNLS) has been obtained by~\citet{trulsen96} for broader bandwidth, only requiring $|\Delta \vec{k}|/k_0\sim \sqrt\varepsilon$, $\varepsilon \ll 1$. 

The effect of currents interacting with an incoming sea state has been studied by~\citet{heller08} using ray dynamics and by~\citet{janssen09} using Monte Carlo simulations. These studies suggest that the interaction between incoming wave and current may serve as a triggering mechanism for the formation of freak waves during nonlinear wave evolution. In the work by~\citet{heller08}, a freak index is defined in terms of the mean wave speed, mean current speed, and the angular spread of the incoming wave, and a quantitative relationship is predicted between this freak index and the occurrence probability of freak waves. However, in this work ray dynamics is used in place of the real wave equation for ocean waves, and nonlinearity is not included in the model. The calculations done by~\citet{janssen09} do include nonlinearity, and serve as a proof of principle for the importance of the interaction between nonlinearity and refraction by currents. Still, results are obtained only for specific values of the input parameters. Thus, a systematic quantitative analysis of the freak wave probability as a function of the strength of nonlinear and refractive effects has not previously been undertaken. 

Here, we use the current-modified nonlinear Schr\"{o}dinger equation (CNLS) developed by~\citet{stocker99} to describe the evolution of the wave envelope, with the aim of integrating both nonlinear effects and wave-current interaction in one model. This approach allows for a full quantitative analysis of both these mechanisms of freak wave formation. In the present paper we focus on numerically exploring the dependence of the wave height distribution on nonlinear effects only, by setting the currents to zero. We will see that interesting behavior is obtained already in the current-free regime. The joint dependence on steepness and current strength for nonzero random currents will be presented in a future publication.

This paper is organized as follows: In section~\ref{sec-model} we review the CNLS equations and present the basic computational setup. In section~\ref{sec-form} we discuss the expected form of the wave height distribution in the presence of deviations from Rayleigh statistics due to nonlinearity and currents. In section~\ref{sec-results} we systematically study the dependence of the wave height distribution on the parameters describing the incoming sea state, allowing the probability of freak wave formation to be calculated as a function of these parameters. Finally, in section~\ref{sec-conc} we summarize our results and discuss the outlook for the future.

\section{Model}
\label{sec-model}
The current-modified nonlinear Schr\"{o}dinger equation equation (CNLS) extends the NLS to include a random surface current varying around a mean value $\vec{u}_0$. The time-independent current field can be expressed as:
\begin{equation}\label{eq:current}
\vec{u}({\vec r})=\vec{u}_0+\delta \vec{u}({\vec r}) \,,
\end{equation}where the random current velocity fluctuations $\delta \vec{u}$
are assumed to be $O(\varepsilon^2)$, and
slowly varying on the scale of a wavelength. 
For surface gravity waves in deep water (i.e., when the water depth is much larger than the wavelength), the dispersion relation is given by:
\begin{equation}\label{eq:dispersion}
\omega (\vec{k},\vec{r})=\sqrt{g|\vec{k}|}+\vec{k} \cdot \vec{u}({\vec r}) \,,
\end{equation}
where $\omega$ is the angular frequency of the incoming wave whose wavelength is given by $\lambda=2\pi/k$, and the wave velocity is $\nabla_k \omega=\sqrt{g/4k} \,\hat{k} +\vec{u}$. For convenience, we take the peak
wave vector of the incoming wave to be in the positive $x$ direction.  The wave vector of each wave component is then expressed as:
\begin{equation}\label{eq:k}
\vec{k}=k_{0}\hat{x}+(dk_x, dk_y) \,, 
\end{equation}
where $k_0>0$.

To simplify the equations and simulations, we work in the frame of reference moving with the peak wave velocity $\sqrt{g/4k_0} \,\hat{x} +\vec{u}_0$, and study the evolution of the wave envelope instead of the wave function itself. The complex wave envelope $a(\vec{r},t)$ is defined by
\begin{equation}\label{eq:wave1}
\zeta(\vec{r},t)={\rm Re} \;a(\vec{r},t) e^{i k_0 x-i \sqrt{g k_0} t} \,, 
\end{equation}
where $\zeta(\vec{r},t)$ is the surface elevation. For a narrow-banded spectrum with small angular spread (i.e., $|\Delta \vec{k}|/k_0\sim \sqrt{\varepsilon} \ll 1$), the wave envelope varies on a scale long compared to the wavelength. In that regime, the probability distribution of the wave envelope $a(\vec r,t)$ closely approximates the crest height distribution. 

Finally, it is convenient to employ dimensionless variables $A=k_0 a$,  $\delta\vec{U}=\sqrt{k_0/g}\delta\vec{u}$,  and
\begin{equation}
(X,Y,Z,T)=(k_0 x-{\sqrt{gk_0}t \over 2},k_0 y,k_0 z,\sqrt{g k_0}t)\,.
\label{dimspacetime}
\end{equation}
In these variables, the third-order CNLS equation is~\citep{stocker99}:
\begin{equation}\label{eq:cnls3}
iA_T-\frac{1}{8}A_{XX}+\frac{1}{4}A_{YY}-\frac{1}{2}A|A|^2-A\delta U_{X}=0\,,
\end{equation}
where subscripts denote partial derivatives.
For larger steepness, i.e., larger $A$, the third-order equation does not sufficiently account for nonlinear effects, and
we need the fourth-order CNLS (CNLS$^4$) equation, also given by~\citet{stocker99}:
\begin{eqnarray}
& & iB _T -\frac{1}{8}(B_{XX}-2B_{YY})-\frac{1}{2}B|B| ^2-B\Phi _{cX}\nonumber\\
& & =\frac{i}{16} (B_{XXX}-6B_{YYX})+\frac{i}{4}B(B\bar{B}^\ast_{X}-6\bar{B}^\ast B_{X})\nonumber\\
& &\;+\bar{\Phi}_{X}B+i(\frac{1}{2}\Phi_{cXT}-\Phi_{cZ})B-i{\nabla}_H\Phi_c\cdot{\nabla}_HB \,.
\label{eq:cnls4}
\end{eqnarray}
On the left hand side of (\ref{eq:cnls4}) are the third-order terms (for comparison with (\ref{eq:cnls3})), while the right hand side contains all terms appearing at fourth order. Here ${\nabla}_H=(\partial/\partial X,\partial/\partial Y)$ is the gradient in the horizontal plane, and $\bar{\Phi}$, $\Phi_c$, and $B$ represent the mean flow, surface current, and oscillatory parts, respectively, of
the velocity potential $\phi$:
\begin{equation}
\phi=\sqrt{\frac{g}{k_0^3}}\left[\bar{\Phi}+\Phi_c+ {\rm Re}\,\left( Be^{k_0z+i\varphi}+B_2e^{2(k_0z+i\varphi)} \right)\right]\,,
\end{equation}
where $\varphi=k_0 x-\sqrt{gk_0}t={X}-T/2$ is the phase. The surface elevation, which is the quantity of interest
for our purposes, is similarly expanded as
\begin{equation}
\zeta=\bar{\zeta}+\zeta_c+{k_0}^{-1}\left[ {\rm Re}\,\left( Ae^{i\varphi}+A_2e^{2i\varphi}+A_3e^{3i\varphi} \right)\right]\,,
\label{Aexpand}
\end{equation}
where the expansion coefficients may be obtained from the velocity potential as 
\begin{eqnarray}
A&=&i B-\frac{1}{2k_0} B_x+\frac{i}{8 k_0^2}(B_{xx}-2B_{yy})+\frac{i}{8} B|B|^2 \nonumber \\
A_2&=&-\frac{1}{2}B^2 +\frac{i}{k_0}BB_x \\
A_3&=&-\frac{3i}{8} B^3 \nonumber \,.
\end{eqnarray}

We have performed calculations using both third-order and fourth-order equations, but all results shown below are obtained using CNLS$^4$. A split-operator Fourier transform method developed by~\citet{weidman86} was employed to solve the PDE numerically. The evolution equation is separated into two parts: a free evolution part and a part containing nonlinear and current terms. In each time step, the wave envelope is transformed into momentum space where the free evolution operator is applied, and then transformed back into position space and acted on by the nonlinear and current operator. In the case of CNLS$^4$, the free evolution part is:
$$
\frac{1}{8}(B_{XX}-2B_{YY})+\frac{i}{16}(B_{XXX}-6B_{YYX})
$$
and the nonlinear and current part is:
\begin{eqnarray}
& &\frac{1}{2}B|B|^2+B\Phi_{cX}+\bar{\Phi}_{X}B+\frac{i}{4}B(B\bar{B}^\ast_{X}-6\bar{B}^\ast B_{X}) \nonumber \\
& &+i(\frac{1}{2}\Phi_{cXT}-\Phi_{cZ})B-i{\nabla}_H\Phi_c\cdot{\nabla}_HB \nonumber \,.
\end{eqnarray}

Without loss of generality, we choose the mean wavelength of the incoming wave to be $156$~m, corresponding to a mean wave speed of $7.8$~m/s. Considering both the computation time and statistical accuracy, the calculation was conducted on a $512$ by $1024$ grid, corresponding to a $10$ km by $20$ km ocean area or $64$ by $128$ wavelengths.

The initial incoming wave is a random linear superposition of ${\cal N}$ monochromatic plane waves with different frequencies and propagating directions:
\begin{equation}\label{eq:iniwave}
\zeta(\vec{r},t)=\frac{1}{\sqrt{\cal N}}
\sum_{i=1}^{\cal N} c_i e^{i\vec{k_i}\cdot\vec{r}-i\omega(|\vec{k_i}|)t} \,.
\end{equation}
Each wave vector $\vec{k_i}$ in (\ref{eq:iniwave}) can be expressed as
\begin{equation}\label{eq:wavenumber}
\vec k=(k_0+k')\cdot (\cos\theta\,\hat x+\sin\theta\,\hat y)\,, 
\end{equation}
where the random wave number variation $k'$ follows a Gaussian distribution of half height width $\Delta k$, and the angle $\theta$ from the mean propagation direction is uniformly distributed in the interval $[-\sqrt{3}\Delta \theta,\sqrt{3}\Delta \theta]$. (To leading order, the results depend only on the standard deviation, $\Delta \theta$, and not on the higher moments of the angular distribution.) Each coefficient $c_i$ is drawn from a normal distribution, and the final sum is normalized to have the desired mean wave height. We have confirmed that our results are ${\cal N}$-independent for sufficiently large ${\cal N}$; specifically, ${\cal N}=2 \cdot 10^{5}$ is used for obtaining the data shown in this paper.

\begin{figure}
  \centerline{\includegraphics[width=\columnwidth]{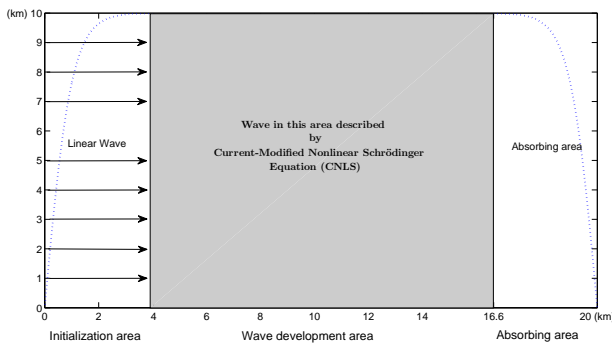}}
  \caption{\label{fig:sketch} \noindent A sketch showing the calculation setup. The $10$~km by $20$~km area is discretized using a $512$ by $1024$ grid. The incoming wave is prepared at the left side of the grid, developing along the $x$ axis from left to right, and is eventually absorbed in the right boundary.}
\end{figure}

The basic setup of the calculation is shown in Figure~\ref{fig:sketch}. The incoming wave is prepared as a random superposition of plane waves at the left side of the grid, developing along the $x$ axis from left to the right, and is eventually absorbed at the right side boundary. The split-operator Fourier transform method used in the calculation automatically imposes periodic boundary conditions on the system. Absorption of the outgoing wave on the right is required to prevent it from re-entering the grid on the left side. However, to avoid reflection or other boundary effects, we allow the outgoing wave on the right to decay gradually using a $\tanh()$ multiplicative factor, and similarly another $\tanh()$ multiplicative factor allows the incoming wave to build up gradually in the initialization area on the left side of the grid, as indicated by the dotted lines in Figure~\ref{fig:sketch}. The boundary conditions in the transverse ($y$) direction remain periodic. By varying the system size in both $x$ and $y$, we have confirmed that for the parameters used below, boundary conditions have a negligible effect on the wave height statistics.

\section{Form of the Wave Height Distribution}
\label{sec-form}
In the Longuet-Higgins random seas model~\citep{longuet57}, the sea state is given by a random superposition of many plane waves with differing directions and frequencies, and by the central limit theorem, the surface elevation function $\zeta({\vec r})$ is distributed as a complex Gaussian random variable with standard deviation $\sigma$. Furthermore, for a narrow-banded spectrum with small angular spread ($|\delta {\vec k}|\ll |{\vec k}|$), the crest height $H$ is equal to the wave function amplitude $|\zeta|$, and the probability distribution of wave height $2H$ is given by the Rayleigh distribution of (\ref{rayleigh}). The corresponding cumulative probability of encountering a wave of height $2H$ or larger is
\begin{equation}
\label{prayleigh}
P_{\rm Rayleigh}(2H) = e^{-(2H)^2/(8 \sigma^2)} \,.
\end{equation}

We now consider the effect of nonlinearity and currents, where the wave envelope $a(\vec r,t)$ defined by (\ref{eq:wave1}) is evolving in accordance with the equations presented in the previous section. Since we still assume a narrow spectrum and a small angular spread, the envelope evolves slowly in space and time, on the scale of the mean wavelength and mean wave period, respectively.
In analogy with the situation for linear ray dynamics in the presence of random currents~\citep{yzhk}, in the neighborhood of any space-time point $({\vec r},t)$, we have a wave intensity proportional to $H^2=|a({\vec r},t)|^2=|\zeta({\vec r},t)|^2$. Thus, in contrast with the original Longuet-Higgins model, the locally-averaged wave intensity is not uniform over all space and time but instead exhibits ``hot spots'' and ``cold spots'' in space-time associated with energy focusing and defocusing. Hence the wave height distribution is given locally in space and time by a Rayleigh distribution around the local mean height (corresponding to a locally random superposition of plane waves), while the local mean height itself varies slowly on the scale of the mean wavelength and mean period. 

At each point in space-time, the central limit theorem and thus the Rayleigh distribution still apply, and we have
\begin{equation}
P_{({\vec r},t)}(2H) = e^{-(2H)^2/(8\sigma^2 I({\vec r},t))} \,,
\label{plocal}
\end{equation}
where $I({\vec r},t)$ is the local energy density, normalized by the average energy density of the incoming sea state (so that $\bar{I}=1$), and
$\sigma^2$ is the variance of the surface elevation in the incoming sea state, before the nonlinear evolution begins to act.

Now averaging over space and over time, we obtain a total cumulative wave height distribution 
\begin{equation}
P_{\rm total}(2H) = \int_0^\infty dI\,  g(I) \, e^{-(2H)^2/(8\sigma^2 I)} \,.
\label{ptotal}
\end{equation}
In (\ref{ptotal}), the full cumulative distribution of wave heights for a given sea state has been expressed as a convolution of two factors: (i) the local density distribution $g(I)$, which at least for small deviations from Rayleigh is conjectured to be $\chi^2$ distributed in analogy to the ray dynamics approximation, and (ii) the universal Longuet-Higgins distribution of wave heights for a given local density. Similar decompositions of chaotic wave function statistics into non-universal and universal components have found broad applicability in quantum chaos, including for example in the theory of scars~\citep{scar1, scar2,baecker}. In the context of rogue waves, a similar approach was adopted by Regev {\it et al.} to study wave statistics in a one-dimensional inhomogeneous sea,
where the inhomogeneity arises from the interaction of an initially homogeneous sea with a 
(deterministic) long swell~\citep{regev}.

Taking the local mean intensity to be $\chi^2$ distributed with $N$ degrees of freedom,
\begin{equation}
g(I)=\chi_N^2(I)=\left(\frac{N}{2}\right)^{\frac{N}{2}}\frac{I^{\frac{N}{2}-1}}{\Gamma\left(\frac{N}{2}\right)}
  e^{-NI/2} \,,
\label{chisq}
\end{equation}
and convolving the $\chi^2$ distribution of the mean intensity with the Rayleigh distribution around the mean intensity, we can obtain as in the linear case~\citep{microwave} a K-distribution for the total distribution of wave heights:
\begin{equation}
P_{\rm total}(H)=2
\frac{\;\;\left({\sqrt{N} H/2\sigma}\right)^{\frac{N}{2}}}{\Gamma(N/2)}
 K_{N/2}\left(\sqrt{N} H \sigma\right)\,,
 \label{kbess}
\end{equation}
where $K_n(y)$ is a modified Bessel function.

Defining the dimensionless variable $x=2H/H_s \approx 2H/(4 \sigma)$, so that a rogue wave is given by $x=2.2$ and an extreme rogue wave by $x=3.0$, we find
the probability of a wave height exceeding $x$ significant wave heights:
\begin{equation}
P_{\rm total}(x)=2
\frac{\;\;\left(\sqrt{N}x\right)^{\frac{N}{2}}}{\Gamma(N/2)}
 K_{N/2}\left(2\sqrt{N}x\right)\,,
 \label{kbess2}
\end{equation}
to be compared with the random seas prediction
\begin{equation}
P_{\rm Rayleigh}(x)=e^{-2 x^2}
\label{prayleighx}
\end{equation} 
in the same dimensionless units.

Notably, the $N$ parameter describing deviations from Rayleigh statistics may be directly related to the excess kurtosis of the sea state, $\gamma_2=\overline{\eta^4}/(\overline{\eta^2})^2-3$, which measures deviations from Gaussian statistics for the surface elevation $\eta$~\citep{onorato02,janssen09}. For a narrow-banded spectrum, we have $\gamma_2=(3/2)\overline{(2H)^4}/\overline{(2H)^2}^2-3$, and from (\ref{kbess}) we obtain 
\begin{equation}
\label{eq:kurtosis}
\gamma_2= \frac{6}{N} \,.
\end{equation}

\section{Numerical Results}
\label{sec-results}

\subsection{Wave Height Probability Distributions}
\label{sub-wh}

In our simulations, we choose the mean wave number to be $k_0=40\;{\rm km}^{-1}$ (a typical value in a normal sea state), corresponding to a group velocity $v_0=7.8\;{\rm m/s}$, wavelength $\lambda_0=156\;{\rm m}$, and period $T_0=10$ seconds. Note that there is no loss of generality in this choice, as it merely sets the fundamental spatial and time scales for the wave dynamics. Each run simulates wave evolution for $4 \cdot 10^5$ seconds, corresponding to $4 \cdot 10^4$ periods. In the setup pictured in Figure~\ref{fig:sketch}, the wave front takes approximately $t=2000$ seconds to travel from the initialization area to the absorbing area taking the straight line path, and the system fully equilibrates at around $t=7000$ seconds. The time interval $7000\;{\rm s}  < t < 400000 \; {\rm s}$ is used to obtain wave height statistics in the wave evolution region defined by $4\;{\rm  km} < x< 16\;{\rm km}$, which indicated by the shaded area in Figure~\ref{fig:sketch}. All wave heights are normalized by the mean wave height $\overline{2H}$ in a given simulation run, obtained using the time interval $7000\;{\rm s}  < t < 20000 \; {\rm s}$. In the end, we calculate the probability distribution of wave heights in each run, and by repeating this process for different input parameters we obtain the dependence of the wave height distribution on the sea parameters.

At present, we set the current to zero, and focus on the nonlinear effects. In this case, we find that the wave height distribution for given sea parameters is reliably obtained from a single run given a sufficiently long run time, as discussed above. The two important properties of the incoming wave are the steepness and the wave vector spread. The steepness $\varepsilon=k_0 \overline{2H}/2$ controls the relative importance of the nonlinear term $\frac{1}{2}B|B|^2$ in (\ref{eq:cnls4}) compared with the linear terms; the steepness is adjusted by varying the mean height $\overline{2H}$ of the incoming sea, since the mean wave number $k_0$ is fixed throughout.

Two parameters are needed to describe the wave vector spread: the variation in magnitude (wave number) is characterized by the ratio  $\Delta k/k_0$ (or equivalently by the frequency spread $\Delta \omega/\omega_0 \approx (1/2)\Delta k/k_0$), and the angular spread is given by $\Delta \theta$. The objective is to find the response of the freak wave occurrence probability to these three input parameters.

First, we keep the wave vector spread constant, and vary the steepness in the range $0.016<\varepsilon<0.07$. The steepness as defined above is normally less than $0.09$ under storm conditions~\citep{dysthe08}. Since we are focused on the normal sea state, ignoring for example tsunamis (large tidal waves caused by underwater earthquake), we are not interested in the behavior at anomalously large values of the steepness. The lower limit is determined by two considerations. Physically, when the steepness becomes very small, the distribution of wave heights approaches the Rayleigh limit, where the wave height distribution is already well understood, and the number of extreme waves is very small.
Numerically, at small values of the steepness, the finite simulation time makes it difficult to quantify deviations from the baseline Rayleigh distribution, due to the paucity of events in the tail of the wave height distribution.

\begin{figure}
\centerline{\includegraphics[width=\columnwidth]{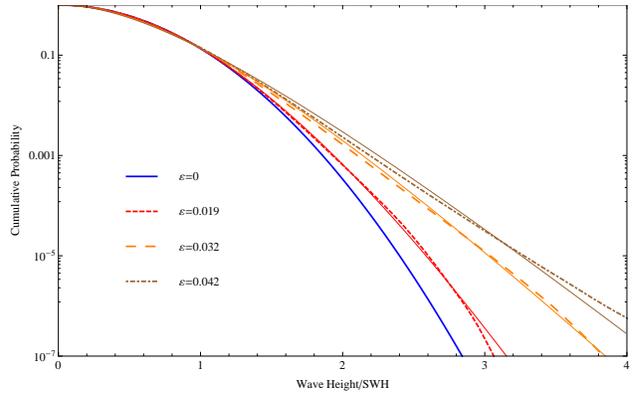}}
\caption{The distribution of wave heights, in units of the significant wave height, is calculated for four nonzero values of the steepness $\varepsilon$ (upper three dashed or dotted curves), and compared with the random seas model of (\ref{prayleighx}) (lowest solid curve). In each case, the solid curve is a best fit to the K-distribution of (\ref{kbess2}). Here the we fix the angular spread $\Delta \theta=2.6^\circ$ and wave number spread $\Delta k/k_0=0.1$ of the incoming sea. 
}
\label{fig:fitdist}
\end{figure}

Typical results are represented by dashed or dotted curves in Figure~\ref{fig:fitdist}, where we fix $\Delta k/k_0=0.1$ and $\Delta \theta=2.6^\circ$ (the values of $\Delta \theta$ required to see very strong effects from nonlinear focusing are typically smaller than those needed to observe significant deviations from Rayleigh by linear scattering~\citep{yzhk}). The cumulative probability distribution of the wave height $2H$, in units of the significant wave height $H_s$, is shown for three nonzero values of the wave steepness $\varepsilon$. From bottom to top, the three thick dashed or dotted lines show results for steepness $\varepsilon=0.019$, $0.032$, and $0.042$. As expected, the Rayleigh probability distribution of (\ref{prayleighx}) is recovered in the limit $\varepsilon \to 0$, and ever stronger enhancement in the tail is observed as the steepness of the incoming sea increases. The occurrence probability of extreme rogue waves, $2H/H_s=3.0$, is enhanced by one to three orders of magnitude for the parameters shown.

The shape of the probability distribution and the typical enhancements in the probability compared with the Rayleigh baseline are consistent with the results of previous numerical simulations. For example, \citet{onorato06} have obtained an enhancement of up to one order of magnitude in the probability of freak wave formation ($2H/H_S>2.2$) and up to $2.5$ orders of magnitude in the probability of extreme freak wave formation 
($2H/H_s>3$), for $\varepsilon=0.08$ ($k_0H_s/2=0.125$), $\Delta k/k_0 = 0.18$ ($\Delta \omega/\omega_0=0.09$), and $\Delta \theta=0$.

In Figure~\ref{fig:fitdist}, each data set is fit to the K-distribution of (\ref{kbess2}), arising from the local Rayleigh approximation. We see that the fits, indicated by solid lines, perform adequately for probabilities down to $10^{-6}$, where statistical noise begins to dominate. The best-fit values are $N=53.2$, $12.4$, and $8.2$, with smaller steepness corresponding to larger $N$. The Rayleigh distribution ($N \to \infty)$ is also shown here, corresponding to $\varepsilon=0$.

In particular, we clearly observe the crossover between the Gaussian behavior (\ref{prayleighx}) at small to moderate heights and  asymptotic exponential behavior at large heights. However, systematic deviations do exist, which are especially visible at larger values of $\varepsilon$, corresponding to smaller values of the $N$ parameter. These systematic deviations are in large part due to the fact that the true wave height distribution for any given set of input parameters exhibits spatial dependence, evolving from the original Rayleigh distribution imposed by incoming boundary conditions to the broader K-distribution, and then gradually back to a Rayleigh distribution as the wave energy is transferred to longer wavelengths and the steepness decreases~\citep{janssen09}. An example of this spatial dependence appears below in Figure~\ref{Fig:spatial}. Thus, a more accurate model for the total wave height distribution consists of a sum of several K-distributions, or equivalently the tail of the full distribution may be modeled by a K-distribution multiplied by a prefactor $C<1$. Nevertheless, as seen in Figure~\ref{fig:fitdist}, (\ref{kbess2}) correctly describes wave height probabilities at the $\pm 20\%$ level of accuracy, allows for an extremely simple one-parameter characterization of the wave height distribution, and facilitates easy comparison between the effects of linear and nonlinear focusing.

\begin{figure}
  \centerline{\includegraphics[width=\columnwidth]{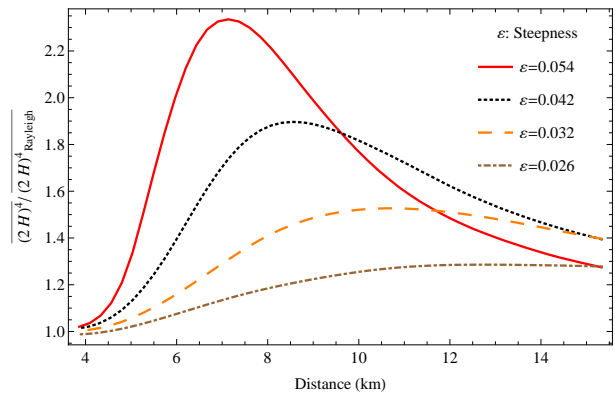}}
  \caption{\label{Fig:spatial} \noindent The fourth moment $\overline{(2H)^4}$ of the wave height distribution, in units where $\overline{2H}=1$, is shown as a function of evolution distance for different values of steepness, starting in each case from a Longuet-Higgins random sea with mean wave speed $v=7.81\,{\rm m/s}$, initial angular spread $\Delta \theta=1^\circ$, and wave number spread $\Delta k/k_0=0.1$. 
  }
\end{figure}

\subsection{Effect of Several Sea Parameters on the Wave Height Distribution}

\begin{figure}
  \centerline{\includegraphics[width=\columnwidth]{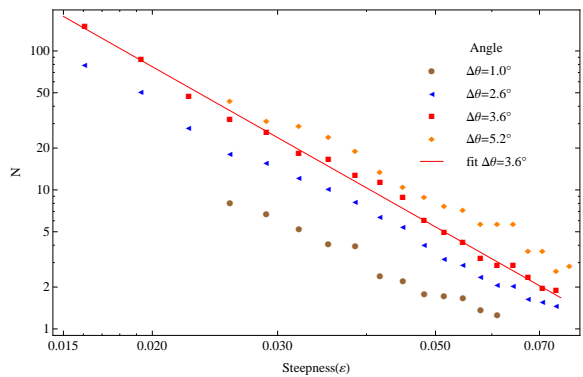}}
  \caption{\label{fig:sn} \noindent  The best-fit $N$ value in (\ref{kbess2}) describing the wave height probability distribution is shown as a function of the steepness $\varepsilon$ for several values of the initial angular spread $\Delta \theta$. The
initial wave number spread is fixed at $\Delta k/k_0=0.1$, as in Figure~\ref{fig:fitdist}. The line shows the best-fit scaling $N \sim \varepsilon^{-2.9}$ for $\Delta \theta=2.6^\circ$.}
\end{figure}

Based on Figure~\ref{fig:fitdist}, we can clearly qualitatively conclude that larger steepness will lead to a higher freak wave occurrence probability. Now by fitting the distribution to (\ref{kbess2}), in Figure~\ref{fig:sn}, we examine quantitatively how increasing the steepness will affect the $N$ value. In the figure, each point corresponds to one particular sea state, and  points of a given shape correspond to the same initial angular spread but different values of the initial steepness. The horizontal axis indicates the steepness, and the vertical axis shows the $N$ value. From bottom to top, the four curves correspond to angular spread $\Delta \theta=1^\circ$, $2.6^\circ$, $3.6^\circ$, and $5.2^\circ$.
A power law is observed on the log-log scale for each angular spread. Taking $\Delta \theta=2.6^\circ$ (indicated by squares in Figure~\ref{fig:sn}) as an example, we obtain the scaling
\begin{equation}
N \sim \varepsilon^c
\label{nonlineps}
\end{equation}
where $c \approx -3$. 

The  $\Delta \theta=3.6^\circ$ and $\Delta \theta=5.2^\circ$ curves lie very close to each other, indeed, at larger values of the steepness (not shown), saturation occurs, indicating that the occurrence rate of freak waves does not fall below a certain level when the steepness is sufficiently high, independent of the angle $\Delta \theta$. However, we also note that the CNLS equations are based on a small steepness approximation, and thus the behavior obtained at very high values of the steepness may not be reliable. In the more realistic low-to-moderate steepness regime shown in the figure, $N$ grows with increasing initial angular spread, i.e., the occurrence rate of freak waves falls. 

Now that we have an understanding of how $N$ changes with steepness, the next step is to fix the steepness and vary the initial angular and wave number spread. At the end, we will be able to bring these results together, to obtain the dependence of the freak wave occurrence rate on all three key environmental parameters.

\begin{figure}
  \centerline{\includegraphics[width=\columnwidth]{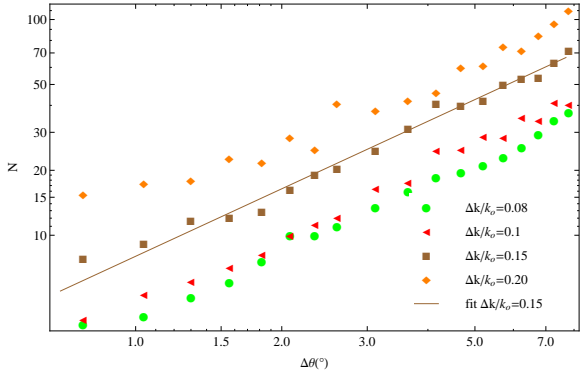}}
  \caption{\label{fig:dn} \noindent The horizontal axis indicates the initial angular spread $\Delta \theta$, and the vertical axis shows the best-fit $N$ value, which determines the wave height probability distribution. Here the steepness is fixed at $\varepsilon=0.032$, and results are shown for four values of the wave number spread $\Delta k/k_0$. The best-fit line shows the scaling of $N$ with $\Delta \theta$ in one example, specifically $N\sim (\Delta \theta)^{1.04}$ for $\Delta k/k_0=0.15$.}
\end{figure}

Figure~\ref{fig:dn} shows the dependence of $N$ on the initial angular spread for several different values of the frequency spread, while the steepness is fixed at $0.032$. In this figure, points of a given shape represent sea states with the same wave number spread. The initial angular spread is varied on the horizontal axis. 

From bottom to top, the four data sets represent wave number spreads $\Delta k/k_0=0.2$, $0.15$, $0.1$, and $0.08$, and the dependence on initial angular spread is well represented by $N\sim (\Delta \theta)^\alpha$, where $\alpha=0.96$, $1.04$, $1.07$, and $1.03$, for the four data sets, respectively. Thus, $N$ scales as the first power of the angular spread at this value of the steepness, independent of the wave number spread.

\begin{figure}
  \centerline{\includegraphics[width=\columnwidth]{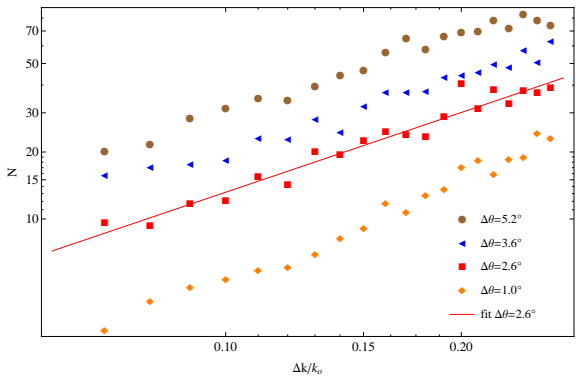}}
  \caption{\label{fig:kn} \noindent The horizontal axis indicates the wave number spread $\Delta k/k_0$, and the vertical axis shows the best-fit $N$ value, which determines the wave height probability distribution.
Again, the steepness is fixed at $\varepsilon=0.032$. The best-fit line shows the scaling of $N$ with $\Delta k/k_0$, in this example, $N \sim (\Delta k/k_0)^{1.15}$ for $\Delta \theta=2.6^\circ$.}
\end{figure}

Similarly, we can choose several typical values of the initial angular spread, and obtain the dependence of $N$ on the initial wave number spread, as shown in Figure~\ref{fig:kn}. Again, the steepness is fixed at $0.032$, and the other physical conditions are the same as in the previous analysis. All sea states having the same initial angular spread are represented by the same symbol. From bottom to top, the angular spread is $\Delta \theta=1^\circ$, $2.6^\circ$, $3.6^\circ$, and $5.2^\circ$. All the initial sea states in this figure have steepness $\varepsilon=0.032$. The power law is consistent for different initial angular spread, showing that $N$ scales approximately as the first power of the wave number (or frequency) dispersion.

We notice deviations from the power-law behavior near the left edge of Figures~\ref{fig:dn} and \ref{fig:kn}, corresponding to very small angular spread $\Delta \theta \approx 1^\circ$ and very small wave number spread $\Delta k/k_0 \approx 0.08$ ($\Delta \omega /\omega_0 \approx 0.04$), respectively. At even smaller values of $\Delta \theta$ and $\Delta k$ (not shown), the CNLS equations we rely on, which
are based on a perturbative expansion in powers of the surface elevation, begin to break down. This breakdown is evidenced, for example by large discrepancies appearing between the wave height distributions obtained using the fourth-order CNLS (\ref{eq:cnls4}) and third-order CNLS (\ref{eq:cnls3}), which make the CNLS expansion untrustworthy.
Furthermore, we recall that the functional form of the K-distribution (\ref{kbess}) is obtained assuming modest fluctuations of the local intensity around the mean intensity, i.e., this functional form makes sense only for $N > 1$, where the wave height distribution may be considered to be a modified Rayleigh distribution. We note in particular that the Benjamin-Feir modulational instability is known to be present in the NLS equations for $(\Delta k/k_0)/\varepsilon <\sqrt{32/\pi}$~\citep{alber}, and for sufficiently small values of $(\Delta k/k_0)/\varepsilon$, this instability will dominate the wave evolution so that extreme waves become commonplace rather than rare events. This regime is outside the range of validity of the present analysis. Nevertheless, as we will see below, the moderate parameter values for which our analysis is applicable, which are also the parameters most likely to occur in nature, can given rise to enhancement factors as large as $10^3$ in the likelihood of occurrence of extreme freak waves (Table~\ref{table:N}).
 
\subsection{A Unified Model}

In the previous subsection, we examined separately the dependence of the wave height distribution on steepness, angular spread, and wave number spread, keeping the other two variables fixed. We would now like to understand the combined dependence on these three parameters describing the initial sea state.

\begin{figure}
  \centerline{\includegraphics[width=\columnwidth]{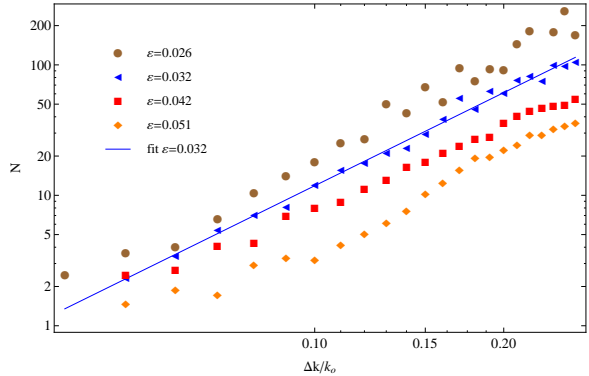}}
  \caption{\label{Fig:kdn} \noindent The best-fit $N$ value in (\ref{kbess2}) describing the wave height probability distribution is shown as a function of wave vector spread (keeping $\Delta k/k_0=7 \cdot \Delta \theta/\pi$ while varying $\Delta k/k_0$ for several values of the steepness $\varepsilon$). The line shows the best-fit scaling with power $2.32$ for $\varepsilon=0.032$.}
\end{figure}

Since both angular and wave number spread control the area in wave vector space in which wave modes are permitted,
we examine a scenario in which the angular spread and wave number spread change at a constant ratio for a given value of the steepness. We arbitrary choose the ratio as $\Delta k/k_0=7 \cdot \Delta \theta/\pi $. The results are shown in Figure~\ref{Fig:kdn}, again plotted on a log-log scale, with the frequency spread $\Delta k/k_0$ on the horizontal axis. Each symbol represents all sea states with a given value of the steepness. The line shows the best-fit power law $N \sim (\Delta k/k_0)^{2.32}$ for $\varepsilon=0.032$. From bottom to top, the other three curves correspond to steepness $\varepsilon=0.051$, $0.042$, and $0.026$, corresponding to a scaling with power  $ 2.3$, $2.0$, and $2.5$, respectively.

First, we focus on the case of steepness $\varepsilon=0.032$, where we have a clear understanding of how $N$ depends on the angular and frequency spread. We notice that the power law exponent $2.32$ appearing in Figure~\ref{Fig:kdn} when $\Delta k/k_0$ and $\Delta \theta$ are varied together is close to the sum of the exponent obtained when $\Delta k/k_0$ and $\Delta \theta$ are varied separately. Thus, we propose a model to describe the dependence of $N$ on angular and frequency spread:
\begin{equation}\label{eq:model1}
N=B \left(\frac{\Delta k}{k_0}\right)^\beta \left(\frac{\Delta \theta}{\pi}\right)^\alpha \,,
\end{equation}
where $B$, $\alpha$, and $\beta$ may depend on the steepness.
Now, if we fix the ratio between the angular and wave number spread, 
$\frac{d\Delta \theta}{\pi}=A \cdot \frac{\Delta k}{k_0}$, 
(\ref{eq:model1}) yields
\begin{equation}\label{eq:model1a}
N=B \left(\frac{\Delta k}{k_0}\right)^\beta  A^\alpha \left(\frac{\Delta k}{k_0}\right)^\alpha=B A^\alpha \left(\frac{\Delta k}{k_0}\right)^{\alpha+\beta} \,.
\end{equation}

Comparing (\ref{eq:model1}) and (\ref{eq:model1a}) with the data shown in Figure~\ref{fig:dn} and Figure~\ref{Fig:kdn}, and setting $A=1/7$, we find the results are consistent and give $B=5279$, $\alpha=1$, and $\beta=1$. Thus, we obtain the following predictive model (when $\varepsilon=0.032$):
\begin{equation}\label{eq:model1b}
N=5300 \cdot \left(\frac{\Delta k}{k_0}\right) \left(\frac{\Delta \theta}{\pi}\right) \,.
\end{equation}

\begin{figure}
  \centerline{\includegraphics[width=\columnwidth]{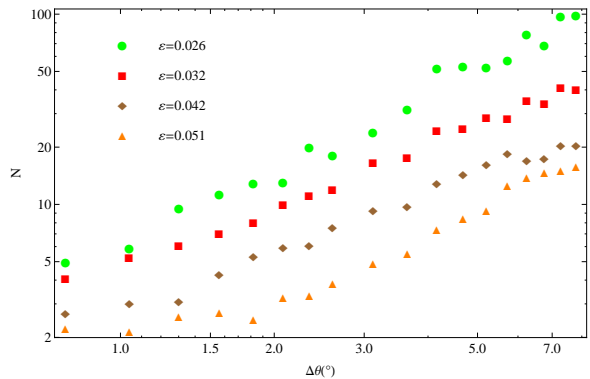}}
  \caption{\label{Fig:dns} \noindent The angular spread $\Delta \theta$ is varied for several values of the steepness $\varepsilon$, keeping the wave number spread $\Delta k/k_0=0.1$ fixed. From bottom to top, the four data sets correspond to $\varepsilon=0.07$, $0.042$, $0.032$, and $0.026$.}
\end{figure}

The next step is to find out how $B$, $\alpha$, and $\beta$ in (\ref{eq:model1}) may evolve when the steepness changes. For several typical values of the steepness $\varepsilon$, we vary the initial angular spread $\Delta \theta$, while keeping the frequency spread fixed at a typical value, $\Delta k/k_0=0.1$.
The results are shown in Figure~\ref{Fig:dns}. Here each symbol represents sea states with one value of the steepness: from bottom to top, the data sets correspond to steepness $\varepsilon=0.051$, $0.042$, $0.032$, and $0.026$. Fitting each data set to a power law relating $N$ and $\Delta \theta$, we find that the power ($\alpha$ in (\ref{eq:model1})) is $0.93$, $1$, $1.15$, and $1.19$ for these four values of $\varepsilon$, respectively. Thus, $\alpha$ is negatively correlated to the steepness, i.e.,
the dependence on angular spread becomes slightly weaker as the steepness increases (consistent with the saturation effect noted earlier). Nevertheless, the effect is quite small for moderate steepness values, so to a good approximation we may treat $\alpha$ as a constant near unity.

\begin{figure}
  \centerline{\includegraphics[width=\columnwidth]{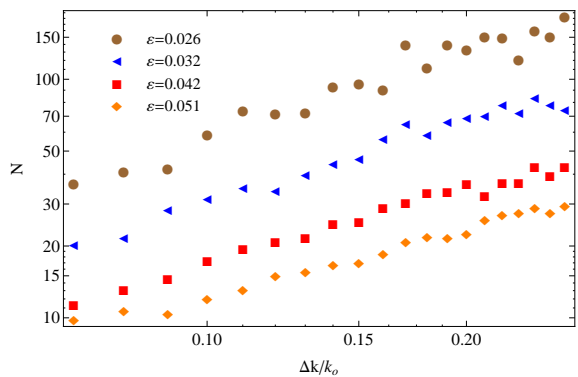}}
  \caption{\label{Fig:kns} \noindent 
  The initial wave number spread $\Delta k/k_0$ is varied for several values of the steepness $\varepsilon$, keeping the initial angular spread $\Delta \theta=5.2^\circ$ fixed. From bottom to top, the four data sets correspond to $\varepsilon=0.07$, $0.042$, $0.032$, and $0.026$.}
\end{figure}

Similarly, in Figure~\ref{Fig:kns} we fix the initial angular spread fixed at a typical value $\Delta \theta=5.2^\circ$, and examine the $N$ dependence on the wave number spread $\Delta k/k_0$ for different values of the steepness $\varepsilon$. 
Each wave height probability distribution is fitted to obtain the $N$ parameter, and a power law relating $N$ and $\Delta k/k_0$ (the $\beta$ exponent in (\ref{eq:model1})) is obtained for each value of $\varepsilon$. From bottom to top, the data sets correspond to steepness $\varepsilon=0.052$, $0.042$, $0.032$, and $0.026$, and yield best fit exponents $\beta=1.07$, $0.98$, $1.03$, and $1.39$, respectively. Since the variation in $\beta$ is non-systematic and weak, we conclude that $\beta$, like $\alpha$, may be approximated as an $\varepsilon$-independent constant near unity.

Thus, we conclude that in (\ref{eq:model1}), both $\alpha$ and $\beta$ are independent of steepness $\varepsilon$, and only $B$ is a function of steepness:
\begin{equation}\label{eq:modelst}
N=B(\varepsilon) \left(\frac{\Delta k}{k_0}\right)  \left(\frac{\Delta \theta}{\pi}\right) \,.
\end{equation}
In Table~\ref{table:model}, we present the value of $B$ corresponding to values of the steepness $\varepsilon$. From Figure~\ref{fig:sn} and (\ref{nonlineps}), we conclude that that $N \sim \varepsilon^{-3}$, If we use this scaling to fit the data in Table~\ref{table:model}, we obtain $B=0.203 \varepsilon ^{-3}$, and thus the final model is described by:
\begin{equation}\label{eq:modelfi}
N=0.203 \; \varepsilon ^{-3}  \left(\frac{\Delta k}{k_0}\right) \left(\frac{\Delta \theta}{\pi}\right) \,.
\end{equation}
As noted previously, the scaling breaks down for very small $\Delta k/k_0$ or $\Delta \theta$, or very large $\varepsilon$, for example, when the two-dimensional Benjamin-Feir index~\citep{mori} is large.

\begin{table}
\caption{Relation Between the Steepness $\varepsilon$ and the Coefficient $B$ in (\ref{eq:modelst})}
\centering
\begin{tabular}{c c}
\hline
Steepness $\varepsilon$ & $\;B$ \\ [1ex]
\hline
0.026 & 17700 \\
0.032 & 5000  \\
0.042 & 1600  \\ 
0.052 & 1100  \\
\hline
\end{tabular}
\label{table:model}
\end{table}

We now examine the implications for the occurrence probability of a freak wave.
According to the conventional definition of a freak wave, the wave height for a freak wave is $3.51\,\overline{2H}$, where $\overline{2H}$ is the mean wave height. By integrating the wave height distribution over heights larger than $3.51\,\overline{2H}$, we obtain the total freak wave occurrence probability. For example, for $\varepsilon=0.032$, $\Delta \theta=5.2^\circ$, $\Delta k/k_0=0.1$, we have $N=21.4$. Comparing the result with a Gaussian theory, the freak wave occurrence probability is 6 times higher. Table~\ref{table:N} below shows the enhancement rate of the probability of freak wave occurrence in the nonlinear theory as compared with the simple Gaussian theory. We note that even at $N$ values between $50$ and $100$, corresponding to the upper range of values in Figures~\ref{fig:sn} through \ref{Fig:kns}, the occurrence of extreme rogue waves is enhanced by an order of magnitude. Exponentially larger enhancement is predicted for parameters associated with smaller values of $N$.

\begin{table}
\caption{Enhancement in the Probability of Rogue Wave Formation\tablenotemark{a}}
\begin{center}
\begin{tabular}{ c  c  c  c }
  \hline
  $\;\;\;\;\;N\;\;\;\;\;$ & $\;\;\;\;\;E(2.2)\;\;\;\;\;$ & $\;\;\;\;\;E(3.0)\;\;\;\;\;$ & $\;\;\;\;\; \gamma_2\;\;\;\;\;$  \\
  \hline
 2 & $1.1 \cdot 10^2$ & $5.2\cdot 10^4$ & 3 \\
  5 & 37 & $7.3\cdot 10^3$ & 1.2 \\
 10 & 16 & $1.3 \cdot 10^3$ & 0.6  \\
 20 & 6.8 & $2.2 \cdot 10^2$ & 0.3 \\
 50 & 2.9 & 27 & 0.12 \\
 100 & 1.8 & 7.8 & 0.06 \\
 \hline
\end{tabular}
\end{center}
\tablenotetext{a}{The enhancement in the probability of rogue wave formation (wave height $2H=2.2\, H_s$) as well as the enhancement of the probability of extreme rogue wave formation (wave height $2H=3.0\, H_s$) and the excess kurtosis $\gamma_2$ (defined in (\ref{eq:kurtosis})) are calculated for several values of the $N$ parameter. Here $E(x)=P_{\rm total}(x)/P_{\rm Rayleigh}(x)$.}
\label{table:N}
\end{table}

\subsection{Spatial Dependence and the Kurtosis}

The distribution data in Figures~\ref{fig:fitdist} to \ref{Fig:kns} is taken over the entire area of wave development thus does not give information about the spatial build up of rogue waves. To study the spatial distribution of the wave heights, we can divide the wave development area into slices of thickness around $0.3$ km (approximately two wavelengths) in the wave propagation direction, and calculate the fourth moment of the wave height distribution, defined as $\overline{(2H)^4}$, in units where $\overline{2H}=1$,
for each slice. For zero steepness, the waves are linear, and the wave height distribution is Rayleigh, giving a value of $3.242$ for the fourth moment of the wave heights. Figure~\ref{Fig:spatial} shows the fourth moment of the wave height as a function of position for different values of the steepness, where the angular spread and wave number variation are fixed at $\Delta \theta=1^\circ$ and $\Delta k/k_0=0.1$.

We can see from Figure~\ref{Fig:spatial} that the fourth moment of the wave height grows initially over a distance scale of several wavelengths, and then returns to the value given by a Rayleigh distribution. The result agrees with the Monte Carlo study conducted by~\citet{janssen09}. 

\begin{figure}
  \centerline{\includegraphics[width=\columnwidth]{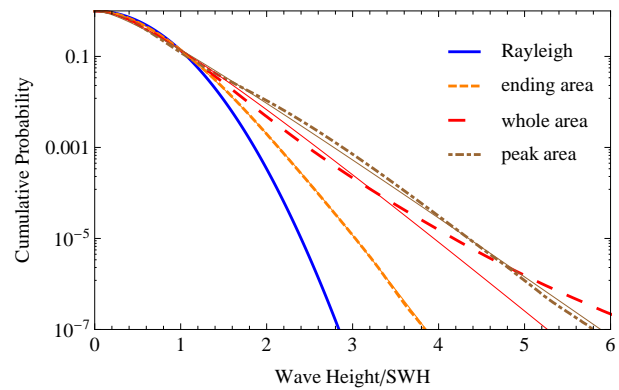}}
  \caption{\label{Fig:slice} \noindent The wave height distribution, in units of the significant wave height, is calculated for three different spatial regions (dashed or dot-dashed lines), and compared with the Rayleigh random seas model of (\ref{prayleighx}) (lowest solid curve). In each case, the solid curve is a best fit to a single K-distribution of (\ref{kbess2}). Here the we fix the angular spread $\Delta \theta=1^\circ$, steepness $\varepsilon=0.054$, and wave number spread $\Delta k/k_0=0.1$ of the incoming sea. 
  }
\end{figure}

In the following, we take a closer look at the wave height distribution in each slice. In this case we choose an example corresponding to the solid red curve in Figure~\ref{Fig:spatial}, with $\Delta \theta=1^\circ$, $\Delta k/k_0=0.1$, and steepness $\varepsilon=0.054$. To reduce statistical noise, we divide the wave development area into larger slices of thickness around $1.3$ km (approximately $8$ wavelengths). The cumulative probability distribution of the wave height $2H$, in units of the significant wave height $H_s$, is shown in Figure~\ref{Fig:slice} for three different areas. The uppermost dot-dashed curve shows the result for the slice centered at the 7 km position; from Figure~\ref{Fig:spatial} we see that this is the peak region for wave height fluctuations. The short-dashed line shows the result for the slice centered at $16$ km, which is near the end of the wave development area, where wave statistics are gradually reverting to the Rayleigh limit. Finally, the long-dashed line in the middle is the wave height distribution for the whole wave development area, as indicated in Figure~\ref{fig:sketch}.

In Figure~\ref{Fig:slice}, each data set is fit to the K-distribution of (\ref{kbess2}), arising from the local Rayleigh approximation. As noted in section~\ref{sub-wh}, when wave height data are collected over a large spatial field that includes some areas of very strong deviations from Rayleigh statistics and other areas where such deviations have not yet had an opportunity to develop, the full distribution may not be well
approximated by a single K-distribution. That is exactly what we observe in Figure~\ref{Fig:slice}: the wave height distribution for a single slice is better approximated by the K-distribution than that for the whole area.

We can also observe that the wave height distribution in a slice in the ending area is better approximated by a single K-distribution than the distribution in the peak area. This is due to the fact that the distribution is evolving much faster in the peak area. Expanding to second order in slice thickness $\Delta x$ for large $N$, the deviation from a K-distribution due to finite slice thickness scales as $(\Delta x/N)^2\cdot (dN/dx)^2$, where  $dN/dx$ is the rate at which N evolves with position in the middle of that slice. This is consistent with the observation in Figure~\ref{Fig:spatial}: the size of each slice is the same, but the slope $dN/dx$ is larger and $N$ is smaller in the peak area. Therefore, the deviation of the wave height distribution from a K-distribution due to finite slice thickness is larger in the peak area than in the ending area.

\begin{figure}
  \centerline{\includegraphics[width=\columnwidth]{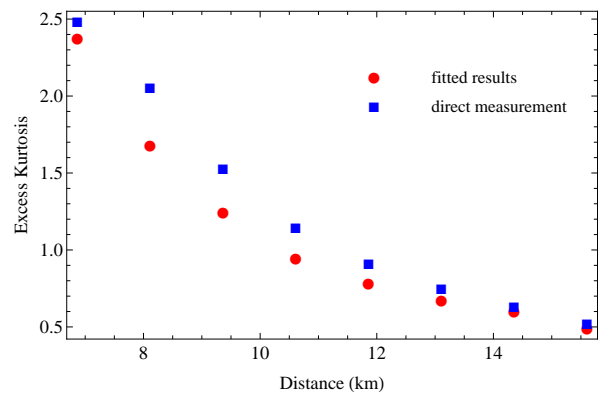}}
  \caption{\label{Fig:ekur} \noindent The excess kurtosis $\gamma_2=(3/2)\overline{(2H)^4}/\overline{(2H)^2}^2-3$ is plotted versus distance along the direction of wave development. The upper data set (squares) is obtained directly from simulation results, and the lower data set(circles) is calculated using (\ref{eq:kurtosis}) from a K-distribution fitted to the wave height distribution in each slice. The parameters are the same as in Figure~\ref{Fig:slice}.
  } 
\end{figure}

In order to further investigate the deviation of the exact wave height distribution from a K-distribution, we compare for each slice the excess kurtosis $\gamma_2$ calculated directly from the simulation with the excess kurtosis obtained by fitting the wave height distribution to a K-distribution and then using (\ref{eq:kurtosis}). Of course, the answers would agree exactly if the distribution within each slice
were perfectly described by a single K-distribution. The results are presented in Figure~\ref{Fig:ekur}, which includes data from $8$ slices, ranging from the peak area ($7$ km) to the ending area ($16$ km). The initial nonlinear focusing area between $4$ km and $7$ km is omitted because the $N$ value is varying too dramatically within a slice, making a fit of the wave height distribution to a single K-distribution meaningless. In the figure, we note that from right to left the difference between the exact kurtosis calculation and the analytic formula of (\ref{eq:kurtosis}) increases systematically, but decreases in the leftmost slice, centered on the peak. If we compare with the solid curve in Figure~\ref{Fig:spatial}, we see that $|dN/dx|$ increases systematically from right to left and then decreases in the peak area. That is consistent with the results in Figure~\ref{Fig:ekur}, given the scaling $\sim(\Delta x/N)^2 (dN/dx)^2$ of the error due to finite slice thickness $\Delta x$.

\section{Conclusions and Outlook}
\label{sec-conc}

In this paper, we study the dependence of the wave height distribution on the steepness, initial angular spread, and frequency of the incoming sea. We use the CNLS$^4$ nonlinear wave equation to simulate the ocean wave development, and obtain quantitative predictions for the wave height distribution as a function of steepness, and angular and frequency spread. We then fit the wave height probability distribution to a K-distribution, governed by a single parameter $N$ that quantifies deviations from the Rayleigh distribution predicted by a simple random seas model. Furthermore, we show that $N$  may be related to the excess kurtosis. We find simple power laws for the dependence of the $N$ parameter on the environment parameters as long as the sea conditions are not extreme, and as a result, we propose a simple model to predict $N$ as a function of the initial sea state. By obtaining $N$ for each sea state, we can quantitatively predict the enhancement of the freak wave occurrence probability for that sea state.

However, this is not a completed investigation. After fully understanding the quantitative consequences of the nonlinear effect, the ultimate goal of the project is to combine nonlinearity and deflection by random currents in a single model, which will allow the probability of rogue wave formation to be predicted for a wide range of realistic sea conditions. Previous linear results show that the effect of current is well characterized by a single parameter: the freak index~\citep{heller08, yzhk}. Preliminary results of combining nonlinearity and deflection by random currents show that the freak index is still a dominant parameter.

When we introduce a nonzero current field, additional length scales come in to play, including the characteristic eddy size (the scale on with the random current varies) and the typical distance required for the first focal point to appear. It is still unclear how these parameters will interact with the wavelength of the incoming wave and the typical distance scale for the buildup of the nonlinear effect. An in-depth investigation is required to understand the underlying mechanism through which the formation of hot and cold spots is aided by nonlinear focusing.

Also, there is a clear need to compare the model simulations with observations and experiments. Although comprehensive global data are not available at this point, it may be possible to compare the results with local observations where data are more readily available, e.g. in the North Sea.


%
%
%
%
%
%
%

\begin{acknowledgments}
This work was supported in part by the US NSF under Grant PHY-0545390. Portions of this research were conducted with high performance computational resources provided by the Louisiana Optical Network Initiative (http://www.loni.org/) and the Center for Computational Science at Tulane University.
\end{acknowledgments}

\end{article}



%

%
%
%
%
%
%
%

\end{document}